\documentstyle[12pt,aasms4]{article}
\begin{document}

\title{Star Formation-Regulated Growth of Black Holes in Protogalactic Spheroids}

\author{Andreas Burkert}
\affil{Max-Planck-Institut f\"ur Astronomie, K\"onigstuhl 17,\\
       D-69117 Heidelberg, \\Germany}
\author{Joseph Silk}
\affil{Department of Astrophysics, University of Oxford, NAPL,\\
Keble Road, Oxford OX13RH,\\ United Kingdom}
\authoremail{burkert@mpia-hd.mpg.de}

\begin{abstract}
 The observed relation  between central black hole  mass and spheroid velocity
 dispersion is interpreted in terms of a self-regulation
model that incorporates a viscous Keplerian accretion disk to feed the black
 hole,
embedded in a massive, self-gravitating star forming disk that eventually
 populates the spheroid. The model leads to a constant ratio
between black
 hole mass and spheroid mass which is equal to the inverse of the critical
Reynolds number for the onset of turbulence in the accretion disk
surrounding the central black hole.
Applying the fundamental plane correlation for spheroids,
we find that the black hole mass 
has a power-law dependence on the  spheroid velocity dispersion
with a slope in the range of 4-5.
We explain the larger scatter in the Magorrian  relation 
with respect to the black hole mass-spheroid velocity dispersion
relationship as a result of secular evolution
of the spheroid that primarily affects its luminosity and to 
a much lesser extent its velocity dispersion.

\end{abstract}


\section{Introduction}

The remarkable dependence of black hole mass on spheroid velocity dispersion,
with low dispersion ($<$10\%) about a mean power-law slope 
between 4 and 5 (Ferrarese \& Merritt 2000, Merritt \& Ferrarese 2000,
Gebhardt et al. 2000a,b) merits
serious attention by theorists. Several scenarios have recently been 
proposed to interpret this relation for either value of the slope,
but none are entirely convincing
(Silk \& Rees 1998, Ostriker 2000, Kauffmann \& Haehnelt 2000, Adams, Graf
\& Richstone 2000).
Silk \& Rees (1998) had predicted such a dependence in a
model that appealed to feedback for quasar outflows on the protogalactic gas
reservoir.  They found the relation 
\begin{equation}
M_{bh}=\sigma_{sph}^5f_{cold}\alpha, 
\end{equation}
\noindent where $f_{cold}$ is the cold gas
fraction and $\alpha$ depends weakly on the presence of an accretion disk
(Haehnelt, Natarajan and Rees 1998) and inversely on the ratio of the
kinetic outflow energy to Eddington luminosity.
This  study requires a very fast accretion phase to feed the 
central black hole.
Recent results by Merrifield, Forbes \& Terlevich (2000) provide
evidence for a relationship between black hole mass and bulge  age
and indicate that the black hole formation timescale is longer,
of order a Gyr.

A flatter relation was derived from cold dark matter-dominated
cosmologies by Kauffmann \& Haehnelt (2000). They assumed that a fixed 
fraction of the cold gas supply that forms spheroid stars in a merger 
will feed the central black hole and discussed the importance of feedback.  
Without feedback their derived scaling relation is 
$M_{bh}\propto\sigma_{sph}^2$. However incorporating feedback leads
to arbitrarily steeper slopes, depending on the choice of the free parameters,
a situation which is not very satisfactory.  A relatively  flat slope
was postulated by Ostriker (2000), for a model in which the central black
hole formed from self-interacting dark matter, now largely discredited
(Gnedin \& Ostriker 2000; Yoshida et al. 2000).  Adams, Graf \& Richstone
(2000) proposed a
new scenario in which the central black hole grew during the protogalactic
collapse  phase by capturing stars on nearly radial orbits.  Their model
is however questionable 
for several reasons: Merrifield's result requires black hole
growth on a time-scale longer than that of protogalactic collapse,
star formation is unlikely to have been highly efficient during this phase, 
and finally they adopt a characteristic velocity dispersion $\sigma$ 
to determine the specific angular momentum of the stars that neglects the 
fundamental plane relation between $\sigma$ and scale radius.

\section{A Model of self-regulated black hole growth}

We propose an alternative model for black hole growth in forming 
spheroids that is based on the commonly adopted  merging scenario
for the formation of  spheroids.
As first proposed by Toomre \& Toomre (1972), spheroids,
like ellipticals, formed via major mergers of gas-rich progenitors.
These disk galaxies most likely had  already formed central black holes 
in their bulges that
satisfied the Magorrian relation (Magorrian et al. 1998).
However the black hole that results
from the coalescence of the progenitor black holes during a spiral-spiral
merger would fall short of the Magorrian relation by a large factor.  A large
fraction of its mass must therefore have been accreted during or after the 
merger event, presumably as gas.

Major merger simulations (Mihos and Hernquist 1996) demonstrate that the gas
rapidly settles, on a
short dynamical time-scale, into a central, self-gravitating disk. By that
time, the
progenitor black holes will also have merged in the centre by dynamical
friction (Burkert \& Sunyaev 2000). The inner disk  has  Keplerian
rotation as the gravitational potential is dominated by the mass of the black halo.
This region looses angular momentum by viscous drag and the
resulting gas inflow will feed the central black halo.
At the same time, the outer self-gravitating disk part is
gravitationally unstable to fragmentation and subsequent star formation.

Within the framework of our model, we can estimate the critical radius $r_{cr}$
that separates the two  regions dominated by accretion and fragmentation
 and which
is determined by the radius out to which the black hole dominates the
gravitational potential,
\begin{equation}
r_{cr}=GM_{bh}/\sigma_{sph}^2.
\end{equation}
Here $\sigma_{sph}$ is the characteristic
velocity dispersion of the spheroid and $M_{bh}$ is the black hole mass.  The
accretion disk is Keplerian with size $r_d$ where it rotates at
\begin{equation}
v_{rot}^2=GM_{bh}r_d^{-1}. 
\end{equation}
Typical time-scales for the gas in the whole inner
disk to accrete onto the black hole are given by the  viscous drag
timescale $t_{vis}=r^2_d/\nu.$ For a viscosity prescription, we adopt the
formulation: $\nu= R_{cr}^{-1}v_{ rot}r_d,$, where  the
critical Reynolds number for the onset of turbulence is $R_{cr}\approx
100-1000$ (Duschl, Strittmatter and Biermann 2000).  The viscous time can now be written as
\begin{equation}
t_{vis}=GM_{bh}R_{cr}\sigma_{sph}^{-3}, 
\end{equation}
\noindent where we have set $r_d=r_{cr}$.

The black hole will grow as long as there is a gas supply for the inner disk.
This gas reservoir is replenished by viscous inflow from the
outer disk into the accretion region and augmented by the increase 
in the inner disk
radius  due to the growth in the black hole mass.
In the absence of any limiting factor all the gas supply will eventually 
be accreted onto the black hole. In this case one would expect a
large variance in the black hole mass with respect to the local velocity
dispersion unless the initial gas fraction of the merger components was
relatively  fine-tuned. This seems unphysical on theoretical grounds,
nor is any relation  observed between gas fraction and velocity dispersion.

We propose that star formation in the outer disk provides
the self-regulation that limits the mass of the central black hole.
Black hole growth  saturates due to the competition with
 star formation, which determines the gas fraction in the disk that is
available
for accretion. 
Let $M_g$ be the total  gas mass available for accretion
and $M_\ast$ be the  mass in stars within the corresponding radius.
We can identify the gas fraction $\epsilon$   as  
$\epsilon = \frac{M_g}{M_d},$ where $M_d=M_g + M_\ast$ is the disk mass 
at a given radius r,
to explicitly demonstrate  how black hole growth is limited by the
 gas supply. Black hole growth only occurs during the gas-rich phase of the
 protospheroid, that is on the  characteristic star formation time-scale
$\tau_{sf}$.

Whatever gas is in the inner disk within $r_{cr}$
will be accreted onto the black hole
without significant star formation, since self-gravity is unimportant.
The  average growth rate of the black hole 
can then be estimated by
\begin{equation}
\dot M_{bh}=\epsilon M_d t_{vis}^{-1},
\end{equation}
\noindent where $M_d$, $\epsilon$ and $t_{vis}$ are defined at  $r_{cr}$.
The total gas fraction within $r_{cr}$ will
contribute to the black hole mass.
We therefore set $M_g = \epsilon M_d$ equal to the black hole mass $M_{bh}$.
This assumes the initial black hole mass to be  negligible
with regard to the accreted mass.
Viscosity (equation 4) then limits  the black hole growth rate to
\begin{equation}
\dot M_{bh}=\sigma_{sph}^{3}R_{cr}^{-1}G^{-1}.
\end{equation}
The growth continues until star formation exhausts the gas supply, on a
time-scale $t_{sf}$, leading to a final black hole mass
\begin{equation}
 M_{bh}=\sigma_{sph}^{3}R_{cr}^{-1}G^{-1}t_{sf}=
1.9\times 10^8\left(\frac{\sigma_{sph}}{200\,\rm km\,s^{-1}}\right)^{3}
\left(\frac{R_{cr}}{1000}\right)^{-1}\left(\frac{t_{sf}}{10^8\rm
yr}\right)\rm M_\odot. 
\end{equation}
This fits the observed normalisation
of the ($M_{bh}$, $\sigma_{sph}$) relation for reasonable values of the star
formation time-scale and the critical Reynolds number for the onset of
turbulence.
 
It is reasonable to assume that molecular clouds form and fragment into
stars on a dynamical
time-scale. We therefore write  the star formation time-scale  as 
$t_{sf}=\eta t_{dyn}=\eta r_e\sigma_{sph}^{-1},$ where $\eta\sim 1.$
Hence 
\begin{equation}
M_{bh}=\eta\sigma_{sph}^{2}R_{cr}^{-1}G^{-1}r_e=M_{sph}\eta
R_{cr}^{-1}.
\end{equation}
This provides a natural explanation of the Magorrian relation.
The ratio of black hole mass to spheroid stellar mass is given by
a ``universal'' constant, $\sim R_{cr}^{-1},$
which in principal should be derivable from fundamental theory and should be
insensitive to  galaxy parameters. It is remarkable that the ratio of two
global masses depends basically on a parameter determined by the microphysics
of viscous transport of angular momentum. The mass ratio could therefore 
provide a direct measure for the typical  Reynolds number $R_{cr}$
in accretion disks
around massive black holes.

The predicted dependence of   
$M_{bh}$ on $\sigma_{sph}$ and $r_e$ follows 
the virial  theorem expectation $M_{sph}=r_e\sigma_{sph}^{2}G^{-1}$
for the spheroid. 
This leads to  a fundamental plane-like projection
on spheroid parameters  for the black hole mass, with a constant offset that is
consistent with the Magorrian relation.
Note that $r_e$ depends on $\sigma_{sph}$ and on surface brightness:
$r_e\propto \sigma_{sph}^2.$ This follows, for example by combining the
Faber-Jackson relation, $M_{sph}\propto \sigma_{sph}^4$ with the virial
theorem. We infer that $M_{bh}\propto \sigma_{sph}^4$, which is in the
observed range. 

The crucial issue, however, is  that of the remarkable reduction  in
dispersion
for the relation between $M_{bh}$ and $\sigma_{sph}$ relative to that
of the spheroid luminosity $L_{sph}$ versus $\sigma_{sph}$.
We note that up to half the scatter in this fundamental plane projection
may be
attributed to  population age differences (Forbes, Ponman and Brown 1999)
that affect primarily the mass-to-light ratios of spheroids.
Another  source of scatter that has also been discussed by
Gebhardt et al. (2000a) is due to projection effects that
lead to variations in 
the effective radius and velocity dispersion 
by a factor of order the typical ellipticity of the
spheroid, up to a factor of 2 (Burkert and Naab 2000).
The inferred intrinsic dispersion  in  the dependence of
bulge mass, and also of
 black hole mass, on velocity dispersion 
must therefore  have been quite small compared to the observed spread.
Note that projection effects would result in scatter in 
the Magorrian relation as well as in scatter in the $M_{bh}$ -- $\sigma_{sph}$
relationship.

The following effect is   likely to play an additional  important role.
Minor mergers add stars but should 
have little effect on the central black hole mass.
These  mergers could occur after the initial protospheroid formation phase, which
characterises the formation of a central disk and the accretion of gas onto
the central black hole but before
the gas in the outer regions of the spheroid was completely consumed.
The mergers induce a change in the star formation rate that is
disproportionately larger than would be inferred directly from
  the amount of stars and gas added.
This is because  the tidal shocks 
stimulate the larger pre-existing outer gas supply
to form stars earlier at an accelerated pace.
Indeed the  NUV-optical color-magnitude relation for early-type cluster
galaxies is incompatible with a monolithic scenario for star formation at
high redshift (Ferreras and Silk 2000).  An increased scatter is found in the
     color-magnitude relation at the faint end, resulting in a significant
     fraction of faint blue early-type systems, implying that  less
     massive galaxies undergo more recent episodes of star formation.
Such episodes will produce scatter in the fundamental plane by affecting the
galaxy  luminosities and outer radii,
but will not 
add appreciably to the scatter in the ($M_{bh}$, $\sigma_{sph}$) correlation
because the velocity dispersion in the inner regions is not changed.

There is a further effect to be explained, namely 
the correlations between fundamental plane residuals (Kormendy 2000),
specifically an anticorrelation between $\Delta\sigma_{sph}$ and
$\Delta r_e.$ 
We attribute this as being in part due to projection effects of
non-axisymmetric spheroids.
In addition late mass loss  or mass infall may well occur.
Substantial  mass loss or  infall  would  modify
both the galaxy size and spheroid velocity dispersion. 
In the limit of an adiabatic response of the host galaxy, we predict that
\begin{equation}
\frac{\Delta r}{r}=-\frac{\Delta \sigma}{\sigma}=-0.4\Delta M_b.
\end{equation}
\noindent which is consistent with the observed anticorrelation beween
$\Delta\sigma_{sph}$ and $\Delta r_e.$
The tightness of the ($M_{bh}$, $\sigma_{sph}$) correlation 
therefore means that late infall into
or outflow from spheroids  can be limited to about 10 percent of the current
stellar mass. This complements a similar conclusion for disks, based on their
observed
thinness (Toth and Ostriker 1992).

Finally  let us introduce
cosmology via merger-induced feeding of the accretion disk.  If the
accretion disk lifetime exceeds the characteristic time between mergers, 
the gas inflow is disrupted.  For example, a binary black hole merger would
sweep out the local environment.  The merger, by forming a transient bar,
will subsequently  resupply cold gas to the central object
causing quasar activity.  Whatever the details, it seems
reasonable to equate the quasar lifetime $t_q = \gamma t_{edd}$
to the merger time-scale $t_{merger}$, thereby
introducing a dependence on the cosmological model. 
Here $\gamma$ (expected to be of order 0.1) is the ratio of $t_q$ to the Eddington time-scale $t_{edd}\equiv 
0.4\rm Gyr.$
 Let $f_{cold}$ be the
cold gas fraction of the total baryon content in the merging galaxy.  One can
then write
\begin{equation}
t_q\equiv t_{merger}=\left(\frac{\bar\sigma_{sph}}{\sigma_0}\right)^{\frac{3(3+n)}{1-n}}
f_{cold}^{\frac{4}{1-n}}  t_0,
\end{equation}
where the normalization assumes
that systems with velocity dispersion
 $\sigma_0=1000\alpha\rm km \,s^{-1}$, with $\alpha\approx 1$,
 are forming at the present epoch $ t_0$.
It follows that
\begin{equation}
M_{bh}=\frac {t_{sf}}{t_q}\left(\frac{\sigma_{sph}}{\sigma_0}
\right)^\frac{3(3+n)}{1-n}
f_{cold}^\frac{4}{1-n}t_0 \frac{\sigma_{sph}^{3}}{GR_{cr}}.
\end{equation}
The expected range for protogalaxies,
$-1.5 \geq n \geq  -2.5,$ again yields the range of 4-5 for the slope of the 
the $M_{bh}$ -- $\sigma_{sph} $  relation.

Note that the 
expected  narrow dispersion in $\sigma_{sph} $ remains 
since most of the dependence on $\sigma_{sph} $ (the part that varies as
$\sigma_{sph}^3 $ ) is cosmology-independent.  Note moreover, that there is
no reliance on the uncertainties inherent in feedback, other than in the
dependence on cold gas fraction, which may be adopted
from the cold gas fraction for the observed damped Lyman alpha clouds.
This varies approximately as $(1+z)^{-3/2}$ for $0<z<3$. We
finally infer an approximately linear dependence of $M_{bh}$ on age of the
stellar population, presumed to have formed when the cold gas fraction was
supplied by the last major merger in accordance with observations.

\section{Discussion}

A  simple model of star formation-regulated black hole growth
seems to explain the observed dependence of black hole on spheroid  mass and
 spheroid velocity dispersion, and especially the observed dispersions.
The proposed model leads to several predictions that might merit
further investigation. We find that the Magorrian relation
depends on two parameters, the critical viscous Reynolds number and the
time-scale of star formation in units of the dynamical time-scale.
As $\eta$ should be of order unity, we infer that black hole growth must be
regulated by 
a turbulent accretion disk with a Reynolds number  of a few hundred.
In addition, the accretion and star formation time-scales are closely
coupled, and in 
our interpretation are of order  the quasar lifetime. 

The scatter of the Magorrian relation directly reflects the scatter in the
fundamental plane parameters $r_e$ and $M_b$, which we postulate to 
be enhanced by 
secular events after black hole and spheroid formation.
Thus the original scatter in the fundamental plane should be as small as that
observed in the 
black hole mass-spheroid velocity dispersion relationship.
This provides a possible observable prediction for high redshift
 field galaxies, and
also for clusters where most of the merging may have been suppressed.

\acknowledgments
We wish to thank UCSC, UC Berkeley, IAP, Chez Papa and  Cafe Peret
for hospitality and for providing a stimulating atmosphere,

\end{document}